\documentclass[aps,twocolumn,superscriptaddress,showpacs,10pt]{revtex4}%
\usepackage{amsfonts}
\usepackage{amsmath}
\usepackage{amssymb}
\usepackage{graphicx}%
\begin{document}
\preprint{ }
\title{Modulational instability of ion-acoustic wave packets in quantum pair-ion plasmas}
\author{A. P. Misra}
\email{apmisra@visva-bharati.ac.in}
\affiliation{Department of Mathematics, Siksha Bhavana, Visva-Bharati university,
Santiniketan-731 235, India.}
\affiliation{Department of Physics, Ume\aa \ University, SE-901 87 Ume\aa , Sweden.}
\author{N. K. Ghosh}
\altaffiliation{Permanent address: Gangapuri Siksha Sadan, Purba Putiary, Kolkata-700 093, India.}

\affiliation{Department of Mathematics, Siksha Bhavana, Visva-Bharati university,
Santiniketan-731 235, India.}

\begin{abstract}
Amplitude modulation of quantum ion-acoustic waves (QIAWs) in a quantum
electron-pair-ion plasma is studied. It is shown that the quantum coupling
parameter $H$ (being the ratio of the plasmonic energy density to the Fermi
energy) is ultimate responsible for the modulational stability of QIAW
packets, without which the wave becomes modulational unstable. New regimes for
the modulational stability (MS) and instability (MI) are obtained in terms of
$\ H$ and the positive to negative ion density ratio $\beta.$ The growth rate
of MI is obtained, the maximum value of which increases with $\beta$ and
decreases with $H$. The results could be important for understanding the
origin of modulated QIAW packets in the environments of dense astrophysical
objects, laboratory negative ion plasmas as well as for the next generation
laser solid density plasma experiments.

\end{abstract}
\received{08 June, 2010}

\revised{16 August, 2010}

\accepted{19 August, 2010}

\startpage{1}
\endpage{102}
\maketitle

In the recent years, there has been a growing interest in investigating
various collective modes and their properties in pair-ion plasmas (see, e.g.,
Refs. \cite{Hasegawa,Cramer,Misra1,Samanta} ). \ Such plasmas are believed to
be ubiquitous in most space and laboratory plasmas \cite{Amemiya,Franklin}.
Moreover, it has been investigated that the pair-ion plasmas have potential
applications in the atmosphere of D-region of Earth's ionosphere, Earth's
mesosphere, the solar atmosphere, as well as in the microelectronic plasma
processing reactors \cite{Kim}. Recent investigations indicate that such
pair-ion plasmas could also be important with regard to the diagnostic point
of view, since the dispersion properties of wave modes can be used to deduce
the plasma parameters \cite{Samanta}. On the other hand, in view of wide
applications in dense astrophysical environments as well as in intense laser
produced plasmas, Misra studied the formation of ion-acoustic shock-like
oscillations in quantum pair-ion plasmas \cite{Misra1}. Again, the propagation
of wave packets in a dispersive nonlinear plasma medium (where the dispersion
is due to quantum tunneling associated with the Bohm potential) has been known
to be subjected to the amplitude modulation, i.e., a slow variation of the
wave packet's envelope due to nonlinearity. The system's evolution is then
governed through the modulational instability (MI). A number of works can be
found in the literature to study the MI in various quantum plasma systems
(e.g., see Refs. \cite{Misra2,Sabry,Misra3}).

In this work, we study the important effect of quantum tunneling associated
with the Bohm potential on the MI of quantum ion-acoustic waves (QIAWs) in an
electron-pair-ion plasma. We show that the quantum coupling parameter $H$ is
ultimate responsible for the stability of modulated QIAW packets, without
which the wave becomes modulational unstable. New regimes for the MI are
obtained with the variation of $H$ along with the positive to negative ion
density ratio $\beta.$ We also obtain the growth rate of MI in terms of the
system parameters.

In what follows, we consider the nonlinear propagation of QIAWs in an
ummagnetized quantum plasma composed of electrons and both positive and
negative ions. The basic normalized equations read \cite{Misra1}

\begin{equation}
\phi=-\frac{1}{2}+\frac{1}{2}{n_{e}^{2/3}}-\frac{H^{2}}{2\mu}\frac{1}%
{\sqrt{n_{e}}}\frac{\partial^{2}\sqrt{n_{e}}}{\partial x^{2}}, \label{e1}%
\end{equation}%
\begin{equation}
\frac{\partial n_{\alpha}}{\partial t}+\frac{\partial(n_{\alpha}u_{\alpha}%
)}{\partial x}=0, \label{e2}%
\end{equation}%
\begin{equation}
\frac{\partial u_{\alpha}}{\partial t}+{u_{\alpha}}\frac{\partial{u_{\alpha}}%
}{\partial x}=-\varsigma_{\alpha}\frac{\partial\phi}{\partial x}, \label{e3}%
\end{equation}%
\begin{equation}
\frac{\partial^{2}\phi}{\partial x^{2}}={(\beta-1)n_{e}}-{\beta}{n_{+}}%
+{n_{-}}, \label{e4}%
\end{equation}
where the suffix $\alpha=+,-$ indicate the quantities for positive and
negative ions respectively. Also, $\phi$ is the electrostatic potential
normalized by $k_{B}T_{Fe}/e$ with $k_{B}$ denoting the Boltzmann constant,
$e$ the elementary charge and $T_{Fe}\equiv\hbar^{2}(3\pi^{2}n_{e0}%
)^{2/3}/2k_{B}m_{e}$ the electron Fermi temperature. Here $\hbar$ is the
Planck's constant divided by $2\pi,$ $m_{e}$ is the electron mass and $n_{e0}$
is its equilibrium number density. Moreover, $n_{\alpha}$ denotes the $\alpha
$-particle perturbed number density normalized by its equilibrium value
$n_{\alpha0},$ $H\equiv\hbar{\omega_{pe}/}k_{B}T_{Fe}$ is the nondimensional
quantum parameter describing the ratio of the plasmonic energy density to the
Fermi energy, $\omega_{p\alpha}\equiv\sqrt{n_{\alpha0}e^{2}/\varepsilon
_{0}m_{\alpha}}$ is the $\alpha$-particle plasma frequency and $m_{\alpha}$ is
the mass. The speed of the $\alpha$-species particle $u_{\alpha}$ is
\ normalized by the quantum ion-acoustic speed $c_{s}\equiv\sqrt{k_{B}%
T_{Fe}/m_{-}},$ the space $(x)$ and time $(t)$ variables are normalized by
$c_{s}/\omega_{p-}$ and $\omega_{p-}^{-1}$ respectively. In Eq. (\ref{e3}),
$\varsigma_{(+,-)}=(m,-1)$ with $m\equiv Z_{+}m_{-}/Z_{-}m_{+}.$ The parameter
$\beta=Z_{+}n_{+0}/Z_{-}n_{-0}$ appearing in Eq. (\ref{e4}) is the positive to
negative ion density ratio with $Z_{+,-}$ denoting the positive (negative) ion
charge states. The second term in the right-hand side of Eq. (\ref{e1}) is due
to the three-dimensional (3D) Fermi-Dirac pressure law for electrons given by
\cite{Landau, Manfredi}%

\begin{equation}
p_{e}=\frac{1}{5}\frac{m_{e}V_{Fe}^{2}}{n_{e0}^{2/3}}n_{e}^{5/3}. \label{e5}%
\end{equation}
Since the equilibrium distribution is always 3D even in one-dimensional (1D)
geometry (we can project the 3D Fermi-Dirac distribution over $x$-direction),
the equilibrium pressure must indeed be given by its 3D expression [Eq.
(\ref{e5})] \cite{Manfredi}.

In order to obtain an evolution equation describing the propagation of
modulated QIAW envelopes we employ the standard reductive perturbation
technique (RPT) in which the independent variables $x$ and $t$ are stretched
as (see, e.g., Ref. \cite{Misra2})$\ \xi$ $=\epsilon(x-v_{g}t)$ and
$\tau=\epsilon^{2}t$, where $v_{g}$ (normalized by $c_{s}$) is the wave's
group velocity\ to be determined by the compatibility condition. The dependent
variables (where the perturbed parts depend on the fast scales via the phase
$kx-\omega t$, and the slow scales only enter the $l$-th harmonic amplitude)
are expanded as%

\begin{align}
\left(  {n_{\alpha},u_{\alpha},\phi}\right)    & =\left(  1,0,0\right)
+\sum_{n=1}^{\infty}\epsilon^{n}\sum_{l=-\infty}^{\infty}\left[  n_{\alpha
l}^{(n)},u_{\alpha l}^{(n)},\phi_{\alpha l}^{(n)}\right]  \nonumber\\
& \times\exp\left[  i(kx-\omega t)l\right]  ,\label{eq6}%
\end{align}
where $\omega,$ $k$ are the normalized wave frequency and wave number
respectively. The state variables $n_{\alpha l}^{(n)}$, etc., satisfy the
reality condition $A_{-l}^{(n)}=A_{l}^{(n)\ast}$ with asterisk denoting the
complex conjugate. Now, substituting the expansion [Eq.(\ref{eq6})] into
equations (\ref{e1})-(\ref{e4}) and expressing the variables $x$ and $t$ in
terms of the stretched coordinates $\xi$, $\tau$, and then collecting the
terms in different power of $\epsilon$ we obtain for $n=1,l=1$ the first order
quantities,
\begin{equation}
n_{e1}^{(1)}=\frac{\phi_{1}^{(1)}}{\Lambda},\text{ }n_{\alpha1}^{(1)}%
=\frac{\varsigma_{\alpha}k^{2}}{\omega^{2}}\phi_{1}^{(1)},\text{ }u_{\alpha
1}^{(1)}=\frac{\varsigma_{\alpha}k}{\omega}\phi_{1}^{(1)},\label{e7}%
\end{equation}
and the following dispersion relation%

\begin{equation}
{\omega^{2}}=\frac{(1+\beta m){k^{2}}}{{k^{2}}+(\beta-1)/\Lambda}, \label{e8}%
\end{equation}
where $\Lambda=1/3+H^{2}k^{2}/4(\beta-1).$ The dispersion relation modified by
the density ratio $\beta$ as well as the quantum parameter $H$ gives two real
eigen modes for the carrier waves. In particular, by disregarding the negative
ion dynamics and considering one-dimensional Fermi pressure law one can
recover the previous result \cite{dispersion}. In the short-wavelength limit
(or for large $k)$, the frequency of QIAWs approaches a constant value
$\sqrt{1+\beta m},$ whereas for large wavelength, the frequency increases with
$k$. The compatibility condition is obtained from the second order $(n=2)$
reduced equations with $l=1$ as
\begin{equation}
{v_{g}}\equiv\frac{\partial\omega}{\partial k}=\frac{\omega}{k(1+{\beta m}%
)}\left[  (1+{\beta m})-{\omega^{2}}+\frac{H^{2}{\omega^{2}}}{{4}\Lambda^{2}%
}\right]  . \label{e9}%
\end{equation}

Proceeding in the same way as in Refs. \cite{Misra2,Misra3} and substituting
all the above derived expressions from $n=2,l=2;n=2,l=0$ into the components
for $n=3,l=1$ of the reduced equations we obtain the following Nonlinear
Schr\"{o}dinger equation (NLSE)%

\begin{equation}
i\frac{\partial\phi}{\partial\tau}+P\frac{\partial^{2}\phi}{\partial\xi^{2}%
}+Q|\phi|^{2}\phi=0, \label{e10}%
\end{equation}
where $\phi\equiv\phi_{1}^{(1)}$ and the dispersive and nonlinear coefficients
$P$, $Q$ are respectively given by%

\begin{align}
{P} &  =\frac{1}{2k(1+\beta m)}\left[  (1+\beta m)\left(  v_{g}-\frac{\omega
}{k}\right)  -\omega^{2}\left(  {3v_{g}}-\frac{\omega}{k}\right)  \right.
\nonumber\\
&  \left.  +\frac{H^{2}\omega^{2}}{4\Lambda^{2}}\left(  3v_{g}-\frac{\omega
}{k\Lambda}\left(  \frac{1}{3}+\frac{5{H^{2}k^{2}}}{{{4(\beta-1)}}}\right)
\right)  \right]  ,\label{e11}%
\end{align}%
\begin{equation}
Q=\frac{\omega^{3}/2k^{2}}{(1+\beta m)}\left[  \frac{1}{\Lambda^{2}}\left[
({{\beta-1)\Delta}}_{1}-{{\Delta}}_{2}\right]  -\frac{k^{2}}{\omega^{2}%
}\left(  \frac{2k}{\omega}{{\Delta}}_{3}+{{\Delta}}_{4}\right)  \right]
,\label{e12}%
\end{equation}
with%
\begin{equation}
{{\Delta}}_{1}=2(A+A_{0})+\frac{1}{9}\left(  \frac{9H^{2}k^{2}}{4\mu
}-5\right)  (F+F_{0}),
\end{equation}%
\begin{equation}
{{\Delta}}_{2}=\frac{1}{4}(2F+F_{0})+\frac{({{\beta-1)}}\left(  7+81H^{2}%
k^{2}/4\mu\right)  }{27\Lambda^{2}},
\end{equation}%
\begin{equation}
{{\Delta}}_{3}=(D+D_{0})\beta m+(E+E_{0}),
\end{equation}%
\begin{equation}
{{\Delta}}_{4}=(B+B_{0})\beta m+(C+C_{0}).
\end{equation}
The coefficients $A,$ $B,...A_{0},$ $B_{0},...$etc., in Eqs. (\ref{e11}) and
(\ref{e12}) are given in the Appendix A. Equation (\ref{e10}) describes the
slow modulation of the first order plasma potential of QIAWs in an
unmagnetized electron-pair-ion plasma. Neglecting the electron dynamics, one
recovers ion-acoustic envelope solitons in a pair-ion plasma \cite{pairplasma}%
. The dispersion coefficient $P\equiv\partial^{2}\omega/2\partial k^{2}$
arising due to quantum diffraction and the charge separation of the plasma
particles, and the nonlinear coefficient $Q$, due to the carrier wave
self-interaction, are significantly modified by the quantum effects, as well
as by the presence of negative ions in our plasma system.

To study the MI of QIAW packets, we assume a monochromatic solution of Eq.
(\ref{e10}) to be of the form $\phi=\phi_{0}$exp$(iQ|\phi_{0}|^{2}\tau)$ (see
e.g., Refs. \cite{Misra2,Misra3}), where $\phi_{0}$ is the constant amplitude
of the carrier wave and $\Delta(\tau)=-Q|\phi_{0}|^{2}$ is the nonlinear
frequency shift. The amplitude and the phase of this solution is modulated
against the linear perturbations as $\phi=[\phi_{0}+\phi_{1}$cos$(K\xi
-\Omega\tau)]$exp$\left[  iQ|\phi_{0}|^{2}\tau+i\theta_{1}cos(K\xi-\Omega
\tau)\right]  $, where $K$ and $\Omega$ are, respectively, the wave number and
the wave frequency of modulation. We then obtain from Eq. (\ref{e10}) the
dispersion relation
\begin{equation}
\Omega=PK^{2}\sqrt{1-\frac{K_{c}^{2}}{K^{2}}}, \label{e13}%
\end{equation}
where $K_{c}=\sqrt{2Q/P}|\Phi_{0}|$ is the critical value of $K$, such that
the MI sets in for $K<K_{c}$, or for wavelengths above the threshold
$\lambda_{c}=2\pi/K_{c}$. The instability growth rate (letting $\Omega
=i\Gamma)$ is obtained as
\begin{equation}
\Gamma=PK^{2}\sqrt{\frac{K_{c}^{2}}{K^{2}}-1,} \label{e14}%
\end{equation}
Clearly, the maximum growth rate obtained at $K=K_{c}/\sqrt{2}$ is
$\Gamma_{max}=|Q||\Phi|^{2}$.\newline

It is evident from Eq. (\ref{e13}) that the instability condition depends only
on the sign of the product $PQ$. Also, relying on the positive or negative
sign of $PQ$ one can predict that when $PQ>0$ the monochromatic waves become
modulationally unstable leading to the formation of a bright soliton, i.e. a
localized pulse-like envelope modulating the carrier wave. When $P/Q<0$, the
solution is modulationally stable, which may result to the formation of a dark
or grey soliton, representing a localized region of decreased amplitude. Thus,
it is reasonable to study the behaviors of $PQ$ , which can be done
numerically with the variation of the system parameters, namely, the density
ratio $\beta$ and the quantum coupling parameter $H$. This is the main purpose
of our present work.

We have considered the density ratio $\beta$ to be greater than unity and the
range of values of $H<1$ for which the electron Fermi speed $v_{F}\equiv
\sqrt{2k_{B}T_{Fe}/m_{e}}$ $<<c$, the speed of light in vacuum and the quantum
collective, mean-field effects become important (where the quantum coupling
parameter $g_{Q}\sim(n_{e0}\lambda_{F}^{3})^{-2/3}\lesssim1,$ where\ $\lambda
_{F}$ describes the scale length for electrostatic screening$)$. The
variations of $P/Q$ with respect to $k$ for different values of $H$ (with
fixed $m$ and $\beta)$ and for different $\beta$ (with fixed $m$ and $H)$ are
shown graphically in Figs. 1 and 2 respectively. When $\beta=4$, i.e., when
the electron concentration is $3/4$ times that of the positive ion, the
stability $(PQ<0)$ and instability $(PQ>0)$ regions (see Fig. 1) for $H=0.75$
and $H=0.7$ are respectively $0<k<0.726;$ $k\gtrsim0.726$ and $0<k<0.4982;$
$k\gtrsim0.4982$. \begin{figure}[ptb]
\begin{center}
\includegraphics[height=3in,width=3.0in]{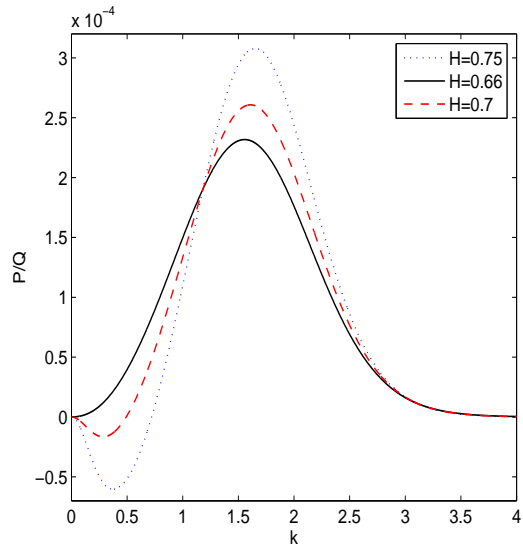}
\end{center}
\caption{(Color online) $P/Q$ is plotted against $k$ to show the stable
$(P/Q<0)$ and unstable $(P/Q>0)$ regions for $\beta=4$, $m=4$ and for
different values of $H$: $H=0.66$ (solid line), $H=0.7$ (dashed line) and
$H=0.75$ (dotted line). The details given in the text.}%
\end{figure}We find from Fig. 1 that for $H\leq0.66,$ $PQ$ becomes positive
for all values of $k$ and any $\beta$. The stability region decreases with
decreasing the values of $H$. Thus, decreasing the values of $H$ or entering
into the density region $n_{e0}\sim1.3490\times10^{31}$m$^{-3}$ or higher, one
can find unstable modulated QIAWs. 

We note that in order to observe different stable and unstable regions for
higher values of $H$ $\geq0.75$ we must have to increase the $\beta$-values as
well to satisfy the restriction for quantum coupling parameter $g_{Q}$ as
described above. Upon increasing the values of the density ratio $\beta$
keeping $H=0.7$ fixed$,$ one can find significant change of the stable as well
as unstable regions. These can be seen from Fig. 2. In this figure, for
$\beta=5,$ $PQ$ is negative in $0<k<0.5755$ and positive otherwise, \ whereas
for $\beta=6,$ $PQ<0$ in $0<k<0.64$ and $PQ>0$ otherwise. Thus, fixing the
electron concentration, and decreasing the positive ion concentration we can
recover the stable regions in a larger domain of $k$. \begin{figure}[ptb]
\begin{center}
\includegraphics[height=3in,width=3.0in]{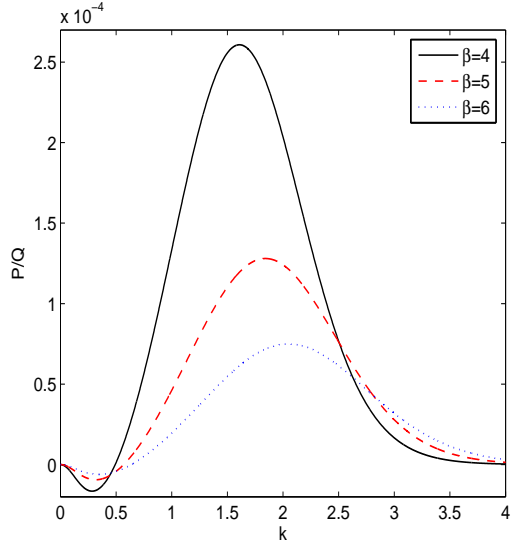}
\end{center}
\caption{(Color online) The same as in Fig. 1, but for $H=0.7$, $m=4$ and for
different values of $\beta$: $\beta=4$ (solid line), $\beta=5$ (dashed line)
and $\beta=6$ (dotted line). The details given in the text.}%
\end{figure}We find that both the dark or grey ($PQ<0$) and the bright
$(PQ>0)$ soliton excitations \ are possible in our quantum plasma system. In
the long-wavelength run $(k<1)$, the former dominates over a long range of
values of $k$ (for a fixed $H$ and increasing $\beta)$, while the latter
exists in a bounded short-range of $k$. 

\ It is to be added that if in any case the plasma system contains negatively
charged dust grains as impurities in the background, one can simply replace
$\beta$ by $\beta-\delta$, where $\delta$ represents the percentage of
negatively charged dust grains with respect to the negative ion concentration.
In presence of negative dusts we may observe that for a fixed $H=0.7,m=4=\beta
,$ $PQ<0$ in $0<k<0.5505$ and positive otherwise for $\delta=0.2,$and for
$\delta=0.4,$ $PQ<0$ in $0<k<0.464$ and positive otherwise. This is how the
charged dust impurity can modify the MI domains. \ Changing $m$ does not
modify the stability or instability regions, but the magnitudes of $P/Q,$
which can be important for calculating soliton widths at different parameters.
The growth rate of MI is also calculated. It is seen that the absolute value
of the maximum growth rate as well as the critical wave number increases
(decreases) with increasing the $\beta$ ($H$) values.

To summarize, we have investigated the instability criteria for the amplitude
modulation of QIAWs in a quantum pair-ion plasma. The quantum coupling
parameter $H$ is shown to play crucial roles in stabilizing the QIAW packets.
New regimes for the MI with the system parameters $H$ and $\beta$ are obtained
in the domain of carrier wave numbers. The results could be important for
negative ion plasmas, forthcoming laser produced plasmas in the laboratory as
well as in dense astrophysical environments.

\textbf{Acknowledgements }

A. P. M. \ gratefully acknowledges support from the Kempe Foundations, Sweden.

\textbf{APPENDIX A}

The coefficients appearing in Eqs. (\ref{e11}) and (\ref{e12}) are given as follows%

\begin{align*}
{A}  & =\left[  \frac{3k^{4}(1-\beta m^{2})}{2\omega^{4}}+{\mu}\left(
{\frac{1}{18}+{\frac{3H^{2}k^{2}}{8\mu}}}\right)  \right]  \\
& \times\left[  \frac{k^{2}(1+{}\beta m-{4\omega^{2}})}{\omega^{2}}-\frac{\mu
}{\left(  1/3+{H^{2}k^{2}/{4\mu}}\right)  }\right]  ^{-1},\text{ }%
\end{align*}

\[
{B}=\frac{mk^{2}}{\omega^{2}}\left(  A+\frac{3mk^{2}}{2\omega^{2}}\right)
,\text{ }{C}=\frac{k^{2}}{\omega}\left(  -A+\frac{3k^{2}}{2\omega^{2}}\right)
,
\]

\[
{D}=\frac{mk}{\omega}\left(  A+\frac{mk^{2}}{2\omega^{2}}\right)  ,\text{ }%
{E}=\frac{k}{\omega}\left(  -A+\frac{k^{2}}{2\omega^{2}}\right)  ,\text{ }%
\]

\[
{F}=\frac{1}{\left(  1/3+H^{2}k^{2}/{\mu}\right)  }\left[  A+\left(  \frac
{1}{18}+\frac{3H^{2}k^{2}}{8\mu}\right)  /\left(  \frac{1}{3}+\frac{H^{2}%
k^{2}}{4\mu}\right)  ^{2}\right]  ,
\]

\bigskip%
\begin{align*}
{A_{0}}  & =\frac{(m^{2}\beta-1){k^{2}}}{v_{g}\omega^{2}}\left(  \frac
{2k}{\omega}+\frac{1}{v_{g}}\right)  -\frac{\left(  9H^{2}k^{2}-32\mu
/3\right)  }{4\left(  1/3+H^{2}k^{2}/4\mu\right)  ^{2}},\\
{B_{0}}  & =\left[  \frac{m^{2}k^{2}}{v_{g}\omega^{2}}\left(  \frac{1}{v_{g}%
}+\frac{2k}{\omega}\right)  +\frac{A_{0}m}{v_{g}^{2}}\right]  ,
\end{align*}
$,$%

\begin{align*}
{C_{0}}  & =\left[  \frac{k^{2}}{v_{g}\omega^{2}}\left(  \frac{1}{v_{g}}%
+\frac{2k}{\omega}\right)  -\frac{A_{0}}{v_{g}^{2}}\right]  ,\text{ }{D_{0}%
}=\left[  \frac{m}{v_{g}}\left(  A_{0}+\frac{mk^{2}}{\omega^{2}}\right)
\right]  ,\text{ }\\
{E_{0}}  & =\left[  \frac{1}{v_{g}}\left(  -A_{0}+\frac{k^{2}}{\omega^{2}%
}\right)  \right]  ,{F_{0}}=\left[  A_{0}+\frac{\left(  9H^{2}k^{2}%
-32\mu/3\right)  }{4\mu\left(  1/3+H^{2}k^{2}/4\mu\right)  ^{2}}\right]  .
\end{align*}

\end{document}